\documentclass[11pt,a4paper]{article}
\usepackage[left=20mm, top=15mm, right=20mm, bottom=25mm, nohead=true, nofoot=true]{geometry}
\usepackage[utf8]{inputenc}
\usepackage{authblk}
\usepackage{indentfirst}
\usepackage{graphicx}
\usepackage{amsmath}
\usepackage{amssymb}
\usepackage{multicol}
\usepackage{bm}
\usepackage{longtable}
\usepackage{pdflscape}
\usepackage{epstopdf}
\usepackage{multirow}
\usepackage{adjustbox}
\usepackage{amsfonts}
\usepackage{ifthen}
\usepackage{fancyhdr}
\usepackage{setspace}
\usepackage{xcolor}
\usepackage[round,authoryear,comma,sort&compress,numbers]{natbib}
\usepackage[toc,title,page]{appendix}
\usepackage[hidelinks]{hyperref}
\usepackage{url}

\hypersetup{
    colorlinks=true,
    linkcolor=green,
    citecolor=blue,
    filecolor=magenta,
    urlcolor=cyan,
    pdftitle={Sharelatex Example},
    pdfpagemode=Null,
}

\fancypagestyle{plain}{\chead{\texttt{\textcolor{brown}{Communications of BAO, Vol. 68, Issue 2, 2021, pp. 447-453}}}}
\title{\textbf{Composition of super-Earths, super-Mercuries, and their host stars}}
\author[1, 2]{V. Adibekyan\thanks{vadibekyan@astro.up.pt, Corresponding author}}
\author[1, 2]{N. C. Santos}
\author[3]{C. Dorn}
\author[2]{S. G. Sousa}
\author[4]{A. A. Hakobyan}
\author[5]{B. Bitsch}
\author[6]{Ch. Mordasini}
\author[1,2]{S. C. C. Barros}
\author[1]{E. Delgado Mena}
\author[1,2]{O. D. S. Demangeon}
\author[1,2]{J. P. Faria}
\author[7,1]{P. Figueira}
\author[1,2]{B. M. T. B. Soares}
\author[8,9]{G. Israelian}

\affil[1]{\scriptsize Instituto de Astrof\'isica e Ci\^encias do Espa\c{c}o, Universidade do Porto, CAUP, Rua das Estrelas, 4150-762 Porto, Portugal}
\affil[2]{\scriptsize Departamento de F\'{\i}sica e Astronomia, Faculdade de Ci\^encias, Universidade do Porto, Rua do Campo Alegre, 4169-007 Porto, Portugal}
\affil[3]{\scriptsize University of Zurich, Institut of Computational Sciences, Winterthurerstrasse 190, CH-8057, Zurich, Switzerland}
\affil[4]{\scriptsize Center for Cosmology and Astrophysics, Alikhanian National Science Laboratory, 2 Alikhanian Brothers Str., 0036 Yerevan, Armenia}
\affil[5]{\scriptsize Max-Planck-Institut f\"{u}r Astronomie, K\"{o}nigstuhl 17, 69117, Heidelberg, Germany}
\affil[6]{\scriptsize Physikalisches Institut, University of Bern, Gesellschaftsstrasse 6, 3012, Bern, Switzerland}
\affil[7]{\scriptsize European Southern Observatory, Alonso de Córdova 3107, Vitacura, Región Metropolitana, Chile}
\affil[8]{\scriptsize Instituto de Astrof\'{i}sica de Canarias, E-38205 La Laguna, Tenerife, Spain}
\affil[9]{\scriptsize Departamento de Astrof\`{i}sica, Universidad de La Laguna, E-38206 La Laguna, Tenerife, Spain}

\def\apj{Astrophys.~J. }

\def\aap{Astron.~Astrophys. }

\def\apjl{Astrophys.~J.~Lett. }

\def\aj{Astron.~J. }
\def\mnras{Mon.~Not.~R.~Astron.~Soc. }

\def\psj{Planet. Sci. J.}

\begin{document}
\pagestyle{empty}
\newpage
\pagestyle{fancy}
\label{firstpage}
\date{}
\maketitle

\begin{abstract}
Because of their common origin, it was assumed that the composition of planet building blocks should, to a first order, correlate with stellar atmospheric composition, especially for refractory elements. In fact, information on the relative abundance of refractory and major rock-forming elements such as Fe, Mg, Si has been commonly used to improve interior estimates for terrestrial planets. Recently \citet{Adibekyan-21} presented evidence of a tight chemical link between rocky planets and their host stars. In this study we add six recently discovered exoplanets to the sample of Adibekyan et al and re-evaluate their findings in light of these new data. We confirm that i) iron-mass fraction of rocky exoplanets correlates (but not a 1:1 relationship) with the composition of their host stars, ii) on average the iron-mass fraction of planets is higher than that of the primordial iron-mass fraction of the protoplanetary disk, iii) super-Mercuries are formed in disks with high iron content. Based on these results we conclude that disk-chemistry and planet formation processes play an important role in the composition, formation, and evolution of super-Earths and super-Mercuries. 
\end{abstract}

\section{Introduction}

The study of only a few giant planets was enough to notice that presence of these planets correlates with stellar metallicity \citep{Gonzalez-97, Santos-01}. Since these pioneering works, different research groups tried to link the chemical composition of stars with the properties of planets \citep{DelgadoMena-10, Adibekyan-13, Adibekyan-15b, Suarez-Andres-17, Brewer-18, Hinkel-19, Teske-19, Adibekyan-19, Unterborn-18}. 

With the increased precision of mass and radius measurements of planets, it become possible to characterize the interiors and bulk composition of low-mass exoplanets \citep{Nettelmann-21, Helled-21}. Several attempts have been made in the last years trying to link the composition of low-mass planets and their host stars. However, these attempts were either based on single planetary systems \citep{Lillo-Box-20, Mortier-20}, on a small sample of planets \citep{Santos-15, Plotnykov-20, Schulze-20}, or on a comparison of the overall properties of planets and overall properties of planet host stars in a population sense \citep{Plotnykov-20}. As a result, it was not possible to reach a firm conclusion either because of low-number statistics or because the results were not as informative (especially if the composition of the stars are not derived in a homogeneous way) as they would be if a direct star-planet comparison was performed.

On the contrary, \citet[][hereafter A21]{Adibekyan-21} adopted a different approach and looked for compositional relation between rocky exoplanets and their host stars. This approach overcomes the following potential issues:

\begin{itemize}
 \item When performing a direct comparison of star-planet compositions for individual systems, large uncertainties in the compositions of planets and/or stars will naturally result in indistinguishable composition of the two.
 \item Because the uncertainties in planetary compositions are typically much larger than those in host star abundances \citep[A21,][]{Schulze-20}, the necessity of precise chemical characterization of planet host stars can be overlooked \citep{Schulze-20}.
 \item Finally, it is impossible to make a general conclusion about the existence of a compositional link between stars and their planets by comparing the compositions of individual planet-star systems.
\end{itemize}

The main findings of A21 can be summarized as follows. The authors selected 22 low-mass exoplanets ($M$ $<$ 10 $M_{\mathrm{\oplus}}$) with precise mass and radius measurements (uncertainty both in mass and radius below 30\%) orbiting around solar-type stars. For the sample planets they determined their normalized planet density ($\rho / \rho_{\mathrm{Earth-like}}$)\footnote{The normalization parameter $\rho_{\mathrm{Earth-like}}$,  is  the density of a planet with Earth-like composition \citep{Dorn17} for a given mass. The normalization is to take into account the dependence of planet density on planet mass for a given composition.} and the iron mass fraction ($f_{\mathrm{iron}}^{\mathrm{planet}}$) using the planet interior models of \cite{Dorn17} and  \citet{Agol-21}. Based on the chemical abundances of the host stars and using the stoichiometric model from \cite{Santos-15,Santos-17} they also estimated the iron-to-silicate mass fraction ($f_{\mathrm{iron}}^{\mathrm{star}}$) of planetary building blocks. Based on these data the authors found: i) the normalized density and iron-mass fraction of exoplanets strongly correlate with the $f_{\mathrm{iron}}^{\mathrm{star}}$; ii) the relation between $f_{\mathrm{iron}}^{\mathrm{planet}}$ and $f_{\mathrm{iron}}^{\mathrm{star}}$ is not 1-to-1 (exoplanets have on average higher $f_{\mathrm{iron}}$ than what is expected from the host star composition), and iii) super-Earths (with $f_{\mathrm{iron}}^{\mathrm{planet}}$ $\lesssim$ 50\%) and super-Mercuries (with $f_{\mathrm{iron}}^{\mathrm{planet}}$ $\gtrsim$ 60\%) appear to be distinct populations in therm of compositions.

\begin{figure}[ht]
\centering
\includegraphics[width=0.85\textwidth]{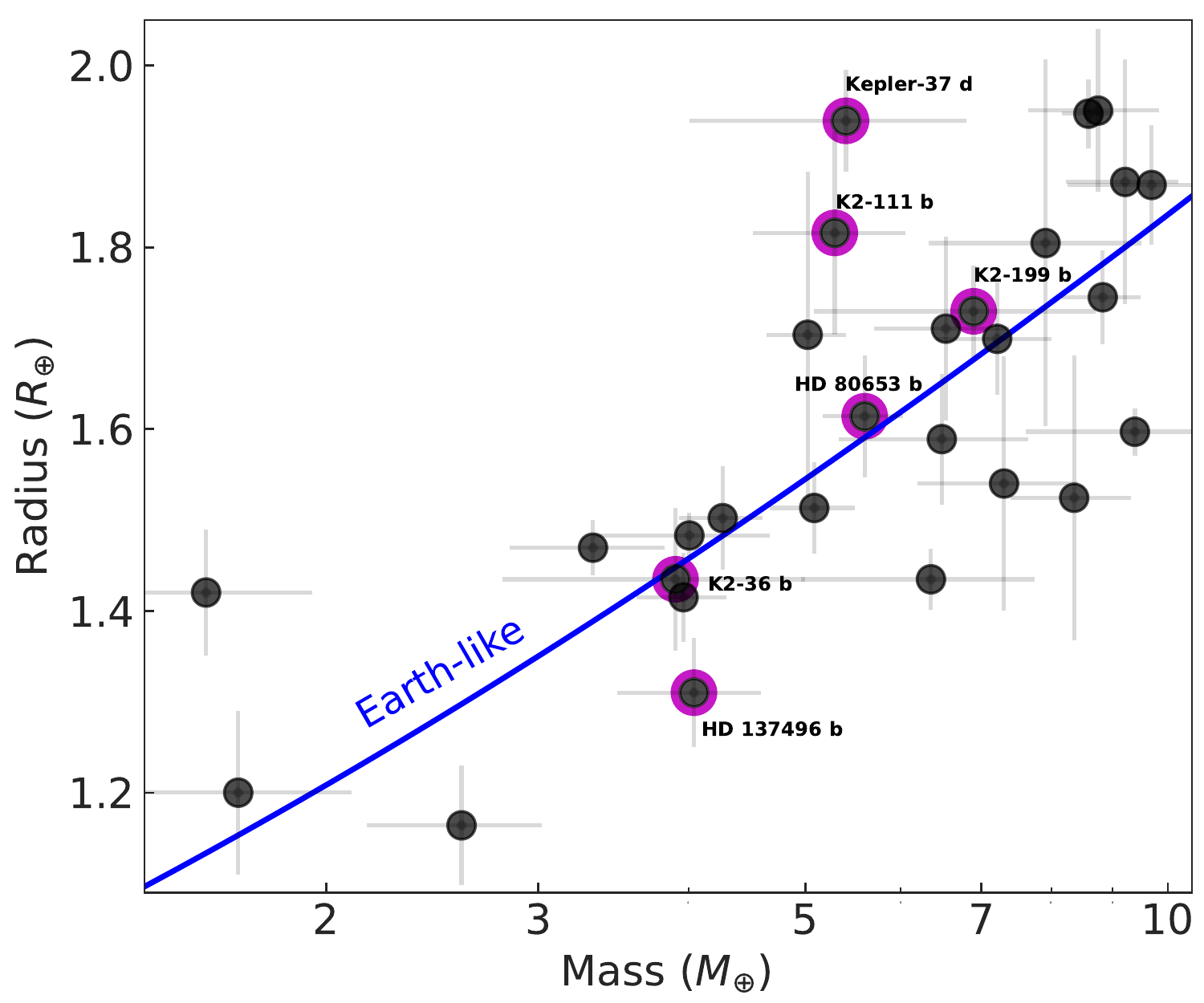}
\caption{Mass-radius diagram for RV-detected planets with masses below 10 $M_{\mathrm{\oplus}}$ and radii below 2 $R_{\mathrm{\oplus}}$ for which the uncertainty both in mass and radius is below 30\%. The blue curve shows the mass-radius relationship for Earth-like composition (32\% Fe + 68\% MgSiO$_{3}$) from \cite{Dorn17}. The six newly added planets are shown in black circles surrounded by a magenta ring. The names of these new planets are also displayed.}
\label{m_r_diagram}
\end{figure}

In this paper, we add six recently discovered exoplanets to the A21 sample and re-evaluate their claims and findings. The distribution of the exoplanets on a mass-radius diagram is shown in Fig.~\ref{m_r_diagram} where we single out the newly added planets.

\section{Properties of planets and their host stars}

To determine the properties of the new planets and their host stars we closely followed the work of A21. Below we provide a brief summary of the methods.

\subsection{Exoplanet properties}

We computed the bulk density ($\rho$) and the normalized density ($\rho / \rho_{\mathrm{Earth-like}}$) of the planets from their mass and radius and the Earth-like composition model of \cite{Dorn17}. From the mass and radius of planets we also estimated their expected iron fraction $f_{\mathrm{iron}}^{\mathrm{planet}}$, which is defined as $(M_{\rm Fe, mantle}+M_{\rm core})/M_{\rm pl}$, where $M_{\rm Fe, mantle}$ and $M_{\rm core}$ are the masses of iron in mantle and core, respectively. For the planet interiors, we assume a pure iron core and a silicate mantle; We neglected possible volatile atmospheric layers. 

\subsection{Host properties}

We used publicly available high-resolution spectra for Kepler-37 \citep[FIES,][]{Telting-14}, K2-36 \citep[HARPS-N,][]{Cosentino-12}, K2-199 \citep[HARPS-N,][]{Cosentino-12}, and HD\,80653 \citep[UVES,][]{Dekker-00} to determine the stellar parameters and abundances of Mg, Si and Fe. The stellar atmospheric parameters ($T_{\mathrm{eff}}$, $\log{g}$, microturbulence (Vmic), and [Fe/H]) of the stars have been determined following the methodology described in our previous works \citep{Sousa-14, Santos-13}. For the derivation of chemical abundances we closely followed the methods described in \citet{Adibekyan-12, Adibekyan-15}. The stellar parameters and abundances of Mg, Si, and Fe of  HD 137496 are taken from \citet{Silva-21} and for K2-111 are taken from \cite{Mortier-20}\footnote{The adopted abundances are determined from the ESPRESSO spectrum.}.

Based on the abundances of Mg, Si, and Fe, and using the stoichiometric models of \citet{Santos-15} we estimated the iron-to-silicate mass fraction ($f_{\mathrm{iron}}^{\mathrm{star}}$) of planetary building blocks under assumption that the stellar atmospheric composition reflects the composition of the proto-stellar (proto-planetary) disk where the star and the planets are formed.

\section{Results}

\begin{figure}[ht]
\centering
  \includegraphics [width=0.85\textwidth] {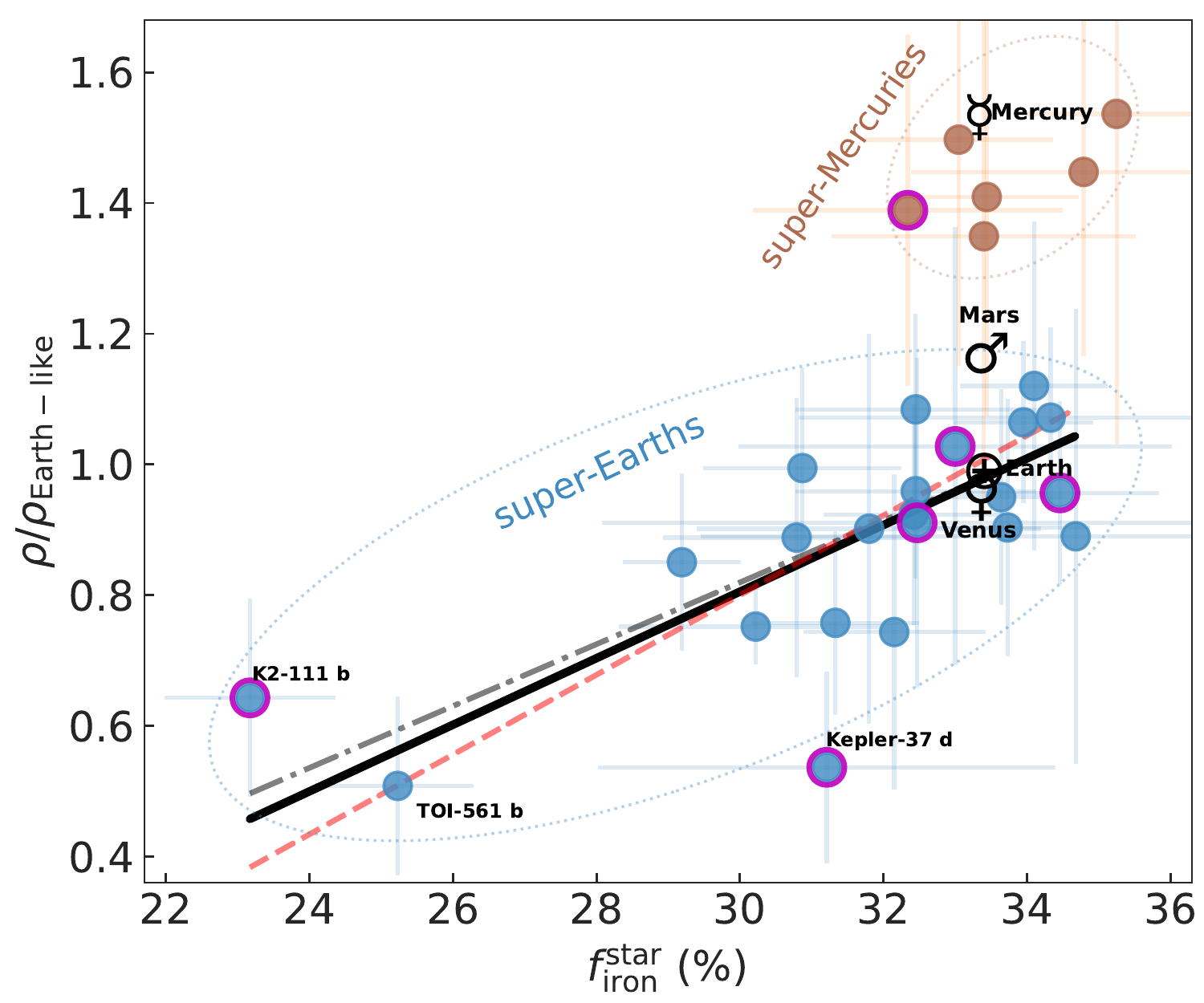}
  \caption{Normalized density of the planets as a function of iron mass fraction of planet building blocks estimated from the host star chemistry. The super-Earths and super-Mercures are shown in blue and brown colors. The positions of K2-111\,b and TOI-561\,b - planets orbiting around metal-poor stars - are indicated. The positions of the Solar System rocky planets are indicated with their respective symbols in black. The red dashed line represent the results of the ODR fit for the super-Earths of the sample of A21. The black solid and dotted-dashed lines show the ODR results for the super-Earths with and without considering Kepler-37 d, respectively. The Solar System planets are not considered in the linear regressions. All error bars show one standard deviation.}
\label{density_firon_star}
\end{figure}

Fig.~\ref{density_firon_star} shows the $\rho / \rho_{\mathrm{Earth-like}}$ as a function of $f_{\mathrm{iron}}^{\mathrm{star}}$.  
The figure reveals a clear correlation between these two quantities indicating that the final planetary density is a function of the composition of the planetary building blocks. We performed an orthogonal distance regression (ODR) and $t$-statistics to quantify the relation and to assess the significance of the relation. The test suggests that the observed correlation is statistically significant with a $p$-value of $\sim$ 3x10$^{-6}$. For the same relation, the sample of A21 revealed a $p$-value of $\sim$ 7x10$^{-6}$. The slopes of the relations obtained for A21 and the extended samples agree withing one-sigma: 0.051$\pm$0.008 vs. 0.061$\pm$0.009.

The largest deviation from the fit is observed for Kepler-37\,d. Kepler-37 is orbited by three\footnote{The presence of a forth, non-transiting planet is unlikely \citep{Rajpaul-21}.} transiting small planets. Kepler-37\,d is the largest planet of the system, the mass of which was very recently determined by \citet{Rajpaul-21} using radial velocity (RV) observations. The authors obtained a RV based mass of 5.4$\pm$1.4 $M_{\mathrm{\oplus}}$ and a dynamical mass of $\sim$ 4 $M_{\mathrm{\oplus}}$. From the low density of the planet, \citet{Rajpaul-21} concluded that either Kepler-37\,d is a water-world ($>$ 25\% H$_{2}$O) or has a gaseous envelope\footnote{Note, that the equilibrium temperature of Kepler-37\,d is about 500K which is the coldest planet in the sample.}. In either case, the planet is most probably not a rocky planet consisting of only metallic core and silicate mantle.  The exclusion of Kepler-37\,d from the ODR slightly reduces the value of the slope (0.047$\pm$0.007) and makes the significance of the relation slightly higher ($p$-value of $\sim$ 1x10$^{-6}$). The results of the ODR fit without considering Kepler-37\,d is shown with a dotted-dashed line in Fig.~\ref{density_firon_star}.

\begin{figure}[ht]
\centering
  \includegraphics [width=0.85\textwidth] {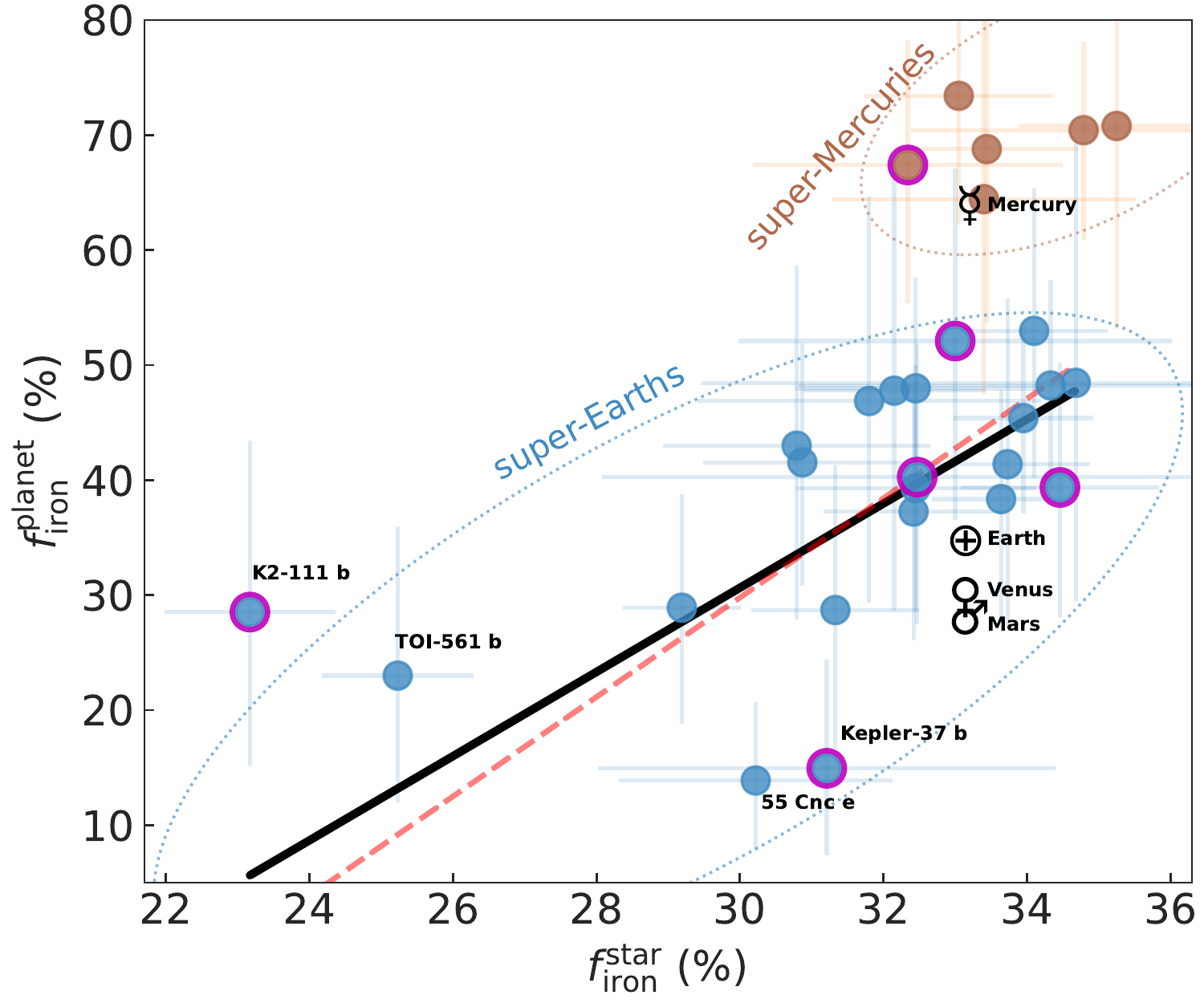}
  \caption{Iron mass fraction of planet building blocks ($f_{\mathrm{iron}}^{\mathrm{star}}$) versus iron mass fraction of the planets ($f_{\mathrm{iron}}^{\mathrm{planet}}$). The meaning of the symbols and lines are the same as in Fig.~\ref{density_firon_star}. The error bars of $f_{\mathrm{iron}}^{\mathrm{star}}$ show one standard deviation. The error bars of $f_{\mathrm{iron}}^{\mathrm{planet}}$ cover the interval between the 16th and the 84th percentiles.}.
\label{firon_planet_star}
\end{figure}

We also study the relation between $f_{\mathrm{iron}}^{\mathrm{planet}}$ and $f_{\mathrm{iron}}^{\mathrm{star}}$ in Fig.~\ref{firon_planet_star}. We performed an ODR and $t$-statistics to the super-Earths and found a $p$-value of $\sim$ 6x10$^{-5}$, which is even smaller than the $p$-value (1x10$^{-4}$) obtained for the A21 super-Earths sample. The slopes of the relations obtained for the A21 and the extended samples agree withing one-sigma: 0.36$\pm$0.9 vs. 4.3$\pm$0.8.

Two planets, 55 Cnc\,e and Kepler-37\,b, show $f_{\mathrm{iron}}^{\mathrm{planet}}$ smaller than $f_{\mathrm{iron}}^{\mathrm{star}}$ of their host stars. We already discussed the possibility for Kepler-37\,b to have a significant amount of volatiles or atmosphere. 55 Cnc multi-planetary system is one of the well studied  ones \citep{Bourrier-18}. Several works suggested that 55 Cnc\,e may have volatile \citep[e.g.][]{Lopez-17} and/or hydrogen \citep[e.g.][]{Hammond-17} layers which could explain the low density of the planet. In addition, it was proposed that 55 Cnc\,e  can have Ca- and Al-rich interior without a significant iron core \citep{Dorn-19}.

\section{Discussion}

It is interesting to see that similarly to TOI-561\,b,  the newly added planet - K2-111\,b,  orbiting a metal-poor star, is also a low-density planet with low iron content. The low $f_{\mathrm{iron}}$ of rocky planets was proposed in \citet{Santos-17} where the authors studied the potential composition of planet building blocks around stars from different Galactic stellar populations. The stoichiometric model of \citet{Santos-15, Santos-17} also suggest a high water-mass-fraction for planets orbiting around metal-poor stars. It is thus possible that both TOI-561\,b and K2-111\,b have a non-negligible volatile layers, which we ignored in our analysis. In a subsequent paper, we plan to model the planet interiors considering also volatile layers and evaluate the presence of correlations between water-mass fraction of the planets and their host stars.

One of the newly added planets, HD 137496\,b, is a super-Mercury. It is intriguing to see that this planet, just like the other five super-Mercuries of A21, has a high $f_{\mathrm{iron}}^{\mathrm{star}}$. A21 suggested that the high iron content of super-Mercuries might be related to the protoplanetary disk composition, and not solely to a giant impact.

\section{Summary}

In this work we extended the sample of \citet{Adibekyan-21} by adding six recently discovered rocky exoplanets and studied the compositional link between rocky exoplanets and their host stars. The main results which confirm the recent findings of A21 are summarized below:

\begin{itemize}
 \item The density ($\rho / \rho_{\mathrm{Earth-like}}$) of super-Earths correlates with the iron content ($f_{\mathrm{iron}}^{\mathrm{star}}$) of the protoplanetary disk. 
 \item There is a non 1-to-1 relation between $f_{\mathrm{iron}}^{\mathrm{planet}}$ and $f_{\mathrm{iron}}^{\mathrm{star}}$. $f_{\mathrm{iron}}^{\mathrm{planet}}$ of super-Earths is larger than the iron content expected from the exoplanet host stars’ composition.
 \item Super-Mercuries are formed in the disks with high $f_{\mathrm{iron}}^{\mathrm{star}}$ suggesting that protoplanetary disk composition might be important for the formation/evolution of these planets. 
\end{itemize}

Studying the relationship between the compositions of planets and their host stars yields a wealth of information on the processes that occur during the formation and evolution of planets. As the number of newly discovered rocky exoplanets continues to increase, we will be able to better understand the origins of these compositional links.


\section*{\small Acknowledgements}
\scriptsize{This work was supported by FCT - Funda\c{c}\~ao para a Ci\^encia e Tecnologia (FCT) through national funds and by FEDER through COMPETE2020 - Programa Operacional Competitividade e Internacionaliza\c{c}\~ao by these grants: UID/FIS/04434/2019; UIDB/04434/2020; UIDP/04434/2020; PTDC/FIS-AST/32113/2017 \& POCI-01-0145-FEDER-032113; PTDC/FIS-AST/28953/2017 \& POCI-01-0145-FEDER-028953. V.A., E.D.M, N.C.S., and S.G.S. also acknowledge the support from FCT through Investigador FCT contracts nr.  IF/00650/2015/CP1273/CT0001, IF/00849/2015/CP1273/CT0003, IF/00169/2012/CP0150/CT0002,   and IF/00028/2014/CP1215/CT0002, respectively, and POPH/FSE (EC) by FEDER funding through the program ``Programa Operacional de Factores de Competitividade - COMPETE''. V.A., E.D.M, N.C.S., and S.G.S. acknowledge support from FCT through Investigador FCT contracts nr.  IF/00650/2015/CP1273/CT0001, IF/00849/2015/CP1273/CT0003, IF/00169/2012/CP0150/CT0002,   and IF/00028/2014/CP1215/CT0002, respectively, and POPH/FSE (EC) by FEDER funding through the program ``Programa Operacional de Factores de Competitividade - COMPETE''. C.D. acknowledges support from the Swiss National Science Foundation under grant PZ00P2\_174028, and the National Center for Competence in Research PlanetS supported by the SNSF. B.B., was supported by the European Research Council (ERC Starting Grant 757448-PAMDORA). O.D.S.D. and  J.P.F. are supported by contracts (DL 57/2016/CP1364/CT0004 and DL57/2016/CP1364/CT0005, respectively) funded by FCT. }

\scriptsize
\bibliographystyle{ComBAO}
\nocite{*}
\bibliography{references}

\newpage
\appendix
\renewcommand{\thesection}{\Alph{section}.\arabic{section}}
\setcounter{section}{0}
\normalsize

\end{document}